\documentclass{rescience}

\usepackage{hyperref}
\usepackage[hyphenbreaks]{breakurl}

\newcommand\hGpc{\mbox{$h^{-1}$\,Gpc}}

\providecommand\aap{Astronomy \& Astrophysics}

\providecommand\jcap{JCAP}

\providecommand\apjs{Astrophys.J.Supp.}

\providecommand\nat{Nature}

\usepackage[numbers]{natbib}

\makeatletter\begin{document}

\begin{tikzpicture}[remember picture, overlay]
  \node [shape=rectangle, draw=darkred, fill=darkred, yshift=-34mm,
        anchor=north west, minimum width=3.75cm, minimum height=10mm]
        at (current page.north west) {};
  \node [text=white, anchor=center, yshift=-39mm, xshift=1.875cm]
  at (current page.north west) {\small
    RESCIENCE C};
\end{tikzpicture}

{\let\newpage\relax\maketitle} \maketitle

\marginnote{
  \footnotesize \sffamily
  \textbf{Edited by}\\
  \ifdefempty{\editorNAME}{\textcolor{darkgray}{(Editor)}}
             {\editorNAME\ifdefempty{\editorORCID}{}{$^{\orcid{\editorORCID}}$}}\\
  ~\\
  \ifdefempty{\reviewerINAME}{}
  {
  \textbf{Reviewed by}\\
  \ifdefempty{\reviewerINAME}{\textcolor{darkgray}{}}
             {\reviewerINAME\ifdefempty{\reviewerIORCID}{}{$^{\orcid{\reviewerIORCID}}$}\\}
  \ifdefempty{\reviewerIINAME}{\textcolor{darkgray}{}}
             {\reviewerIINAME\ifdefempty{\reviewerIIORCID}{}{$^{\orcid{\reviewerIIORCID}}$}\\}
  ~\\
  }
  \textbf{Received}\\
  \ifdefempty{\dateRECEIVED}{---}{\dateRECEIVED}\\
  ~\\
  \textbf{Published}\\
  \ifdefempty{\datePUBLISHED}{---}{\datePUBLISHED}\\
  ~\\
  \textbf{DOI}\\
  \ifdefempty{\articleDOI}{---}{\articleDOI}
}

\newcommand{\container}[1]{\def\@container{#1}}
\begin{container}
  \afterpage {
    \begin{statement}
      \scriptsize \sffamily
      \hrule \vskip .5em
      Copyright {\textcopyright} \articleYEAR~\authorsABBRV,
      released under a Creative Commons Attribution 4.0 International license.

      Correspondence should be addressed to
      \contactNAME~(\href{mailto:\contactEMAIL}{\contactEMAIL})

      The authors have declared that no competing interests exist.

      \ifdefempty{\codeURL}{}
      {Code is available at
      \href{\codeURL}{\detokenize\expandafter{\codeURL}}\ifdefempty{\codeDOI}{.}{ -- DOI \doi{\codeDOI}.}\ifdefempty{\codeSWH}{.}{ -- SWH  \href{https://archive.softwareheritage.org/\codeSWH/}{\detokenize\expandafter{\codeSWH}}.}}

      \ifdefempty{\dataURL}{}
      {Data is available at
      \href{\dataURL}{\detokenize\expandafter{\dataURL}}\ifdefempty{\dataDOI}{.}{ -- DOI \doi{\dataDOI}.}}

      \ifdefempty{\reviewURL}{}
     {Open peer review is available at \href{\reviewURL}{\detokenize\expandafter{\reviewURL}}.}
    \end{statement}
  }
\end{container}

\abstract{\articleABSTRACT} 
\cite{RBG08}

\section{Introduction}

This paper studies the reproducibility of the main observational
results of a cosmic topology research paper published by myself and
co-authors in 2008\cite{RBG08}. The paper used the
surface-of-last-scattering optimal cross-correlation method of finding
a preferred orientation of the fundamental domain of the spatial
section of the Universe, under the working hypothesis that the spatial
section is a Poincar{\'e} dodecahedral space\cite{LumNat03}. The
code was developed by me, with comments provided by my coauthors. The
results that should be reproduced are those that use the method
described in Section 3.2 of RBG08, and the observational analysis
results described in Section 4.2, displayed in Figs.~3, 4, 5 and given
numerically in Tables 2 and 3 of RBG08. Related cosmic topology papers
by other authors are published with no references to software package
details or software licences.

The reason for attempting and documenting the reproducibility of this
paper is that not only are many papers in astronomy\cite{Allen18}
and other fields still published without providing the full
empirical data sets and source code under free-software licences,
but even those that provide free-licensed software and input data
may be difficult to reproduce\cite{Ioannidis2009,Chang15,Stodden18}.
While observational data in cosmology are usually made available online
with high-quality documentation, often after an embargo period,
free-licensed software in the field of cosmic topology,
in particular, library functions
for defining matched circles in the cosmic microwave background
or matched \emph{discs} in extragalactic 3-dimensional comoving space\cite{RK11},
is only recorded in the scientific literature in papers published by my research group.

To document and help analyse the success and difficulties in reproducing
scientific results in this context,
the editors of {\it ReScience C} posed the ``Ten Years Reproducibility
Challenge'', a request that scientists attempt to
reproduce the main results of \emph{their own}
peer-reviewed scientific research papers that had been
published before 1 January 2010, and document the method and results
in {\it ReScience C}\cite{TenYrChallenge20}.

\section{Method}

The first steps for trying to reproduce the original results
of RBG08 were to (re-)read the appropriate sections of the paper,
initially taking the view of a non-author.
\begin{enumerate}
\item
  Section 2.1\cite{RBG08} states that the analysis method of
  Section 3.2 requires the three files at URLs listed in footnotes 1,
  2, 3 on the same page. These files represent two versions of an
  all-sky map of the Universe mostly representing cosmic microwave
  background emission at 10{\hGpc} (comoving) from the Earth as
  observed by the Wilkinson Microwave Anisotropy Probe
  (WMAP)\cite{WMAP5Hinshaw}, and the ``kp2'' mask to enable
  analysis that avoids the most contaminated regions of the sky. These
  files need to be downloaded.
\item
  Footnote 7\cite{RBG08} indicates that {\sc circles-0.3.2.1}, to
  be found at the URL \url{http://cosmo.torun.pl/GPLdownload/dodec/},
  provides the software for generating the figures and tables. This
  software needs to downloaded from
  \url{http://cosmo.torun.pl/GPLdownload/dodec/circles-0.3.2.1.tar.gz}.
\end{enumerate}

The next step was to develop a script on a {\sc git} repository server
that satisfies the requirements of the international scientific
community, specifically the International Science
Council\cite{ISCFreedoms}, by not blocking access to scientists
of any countries or territories.  During 2018 and 2019, several of the
most popular {\sc git} repository servers partially blocked access to scientists
and other residents of several countries and territories
(Github\cite{Github2020ToS,TechCrunch2019github,Zdnet2019github},
Bitbucket\cite{Atlassian2019bitbucket},
Gitlab\cite{Gitlab2018ban}; the {\sc gitlab} software is free-licensed
and can be installed independently of the Gitlab online service).
The bans have presumably continued into 2020.
A shift of my own software to servers acceptable under international
scientific ethical standards is underway, but incomplete as of early
2020. I chose a community-based server, {\em Codeberg}, not currently listed
on the Wikipedia list of source code hosting
facilities\footnote{\url{https://en.wikipedia.org/wiki/Comparison_of_source-code-hosting_facilities}}.
In 2019, the {\em Investigating \& Archiving the Scholarly Git Experience} project team
expressed its concerns about the bans, describing them as having ``far-reaching and chilling consequences
for open source, open scholarship, and for the open exchange of information and ideas''\cite{IASGE2019}.

The remaining planned steps were to implement the minimal number of
updates to make the code work and replicate the original results,
using modern hardware and a modern software environment. Footnote 7 of
RBG08 warns that ``These [{\sc circles-0.3.2.1} and {\sc
    circles-0.3.8}] and earlier versions of the software require
medium to advanced {\sc GNU/Linux}, {\sc Fortran77} and {\sc C}
experience for a scientific user.'' There is no statement regarding
the particular compiler(s) used.  As far as I recall, it's very likely
that the widely used GNU fortran compiler of the time, {\sc g77}, was
used together with {\sc gcc}, as selected automatically by {\sc
  autotool} packages.

The system and hardware chosen for the reproduction project were an
{\sc AMD} computer running with a {\sc Debian GNU/Linux 9.12} system on an
{\sc x86\_64 Linux-4.9.0 kernel}.
The Fortran compiler chosen was {\sc GNU Fortran (Debian
  6.3.0-18+deb9u1) 6.3.0 20170516}.

\begin{table}
  \begin{tabular}{ll}
    \hline
    ded213c1c4cfdfe2ef92f7155b27d58c & wmap\_ilc\_5yr\_v3.fits \\
    fbc8b2518fdddf0a1e7b5acde99a748e & wiener5yr\_map.fits \\
    5aa3267dc6d69bf8c5f0a3a893e23960 & wmap\_kp2\_r9\_mask\_3yr\_v2.fits \\
    afbd67d8120c11e949eb0c414c2775f5 & circles-0.3.2.1.tar.gz \\
    \hline
  \end{tabular}
  \caption{Checksums (md5sums) of the data and the main software
    source code files of RBG08, use for the present reproducibility
    test.\protect\label{t-md5sums}}
\end{table}

\section{Results}

The overall script intended to carry out the full sequence of
downloads, configuring of packages, compiling of packages,
subdirectory user-level installation of packages, setting up of
calculation parameters, and running the main code, was set up as a
{\sc bash} script {\tt reproduce\_RBG08.sh}.

The full package aiming to reproduce the figures and tables listed
above is provided at \url{https://codeberg.org/boud/0807.4260},
named after the ArXiv identity of RBG08.

\subsection{Downloading data and software source code}

\begin{enumerate}
\item
  The URL in footnotes 1, 2 and 3\cite{RBG08} gave clickable
  links that were split into two and not correctly clickable. The user
  needs to cut/paste the two halves of each URL in order to obtain the
  three data files. The data files were downloaded with no apparent
  problem, with md5sums as indicated in Table~\ref{t-md5sums}.
\item
  The file {\sc circles-0.3.2.1.tar.gz} with the md5sum indicated in
  Table~\ref{t-md5sums} was downloaded. It was included in
  the main git repository in its original form. Subsequent changes
  are recorded in the git history at
  \url{https://codeberg.org/boud/0807.4260}.
\end{enumerate}

\subsection{Compiling/debugging}

Fixes needed in order to successfully compile {\sc circles} include:
\begin{enumerate}
\item
  A Fortran 77 line that ended on one line with a $+$ symbol and
  started on the next line with another $+$ symbol (within the valid
  columns for standard Fortran 77) was apparently accepted by the {\sc
    gcc} family fortran compiler in 2008, but not now (2020). One of
  the $+$ symbols was removed.
\item
  A fitting algorithm {\sc gsl\_multifit\_covar} available in GNU
  Scientific Library ({\sc GSL}) versions 1.x was obsoleted; it is no
  longer present in modern 2.x versions of {\sc GSL}. With the aim of
  minimising the interventions required in the system, {\sc
    GSL-1.10} was downloaded and compiled from source, re-creating
  part of the original software environment.
\item
  Using the modern {\sc gfortran} compiler options {\tt -fcheck=bounds
    -Wall} to highlight likely sources of bugs due to insufficiently
  standard coding led to many warnings.  Checking of these warnings
  motivated many fixes that could be expected to either solve errors
  in running the main {\sc circles} package, or reduce the chance of
  calculational errors.
\item
  The autotools {\tt autoreconf} command was run in the main {\sc
    circles-0.3.2.1} directory and its subdirectories.
\item
  The {\sc cosmdist} package provided by default in a subdirectory of
  {\sc circles-0.3.2.1} was replaced by a
  download/configure/compile/install section of the main reproduction
  script, since {\sc cosmdist} is now available in an online
  git repository. The aim was to reduce the chance of {\sc cosmdist}
  being a blocking factor in reproduction of the calculations.
\item
  Memory allocation errors that occur for running {\sc circles}
  without previously definining environment variables for key
  values such as input filenames are present in {\sc
    circles-0.3.2.1}. These were most likely not noticed in RBG08
  because of the use of environment variables providing these
  values. Fixes to the front-end C file {\sc circles.c} were
  made with the intention of avoiding memory allocation errors,
  which are typically reported to the user as segmentation faults.
  However, memory allocation errors remained, most likely due
  to Fortran 77--C interfacing issues. This is described in the
  \S\ref{s-running} on attempted running of the code.
\end{enumerate}

\subsection{Dependencies}

The full list of dependencies listed -- not necessarily really used -- in
the final {\sc gfortran} compile operation that creates the {\sc circles}
binary executable is:
{\tt \mbox{\tt -lpgplot} -lpng lcosmdist -lisolat -lastromisc -llapack -lcblas
-lf77blas -latlas -lgfortran -lquadmath -lcfitsio -lcosmdist -lgsl
-lgslcblas -lm -lgcc -lX11 -lm}. \sloppy

\subsection{Running and a software evolution block} \label{s-running}

\subsubsection{Front end user-friendliness}

At the time of writing the original paper, the aim was that the use of GNU
tools to provide a free-licensed package configurable and compilable
with {\tt ./configure \&\& make} and a detailed {\tt ./circles
  -{}-help} command would be sufficient to enable easy reproduction by
a scientifically competent user. For example, invoking the {\sc
  circles} help option was intended to show both single-hyphen, one-character options
and their equivalent double-hyphen, long options, such as
\mbox{{\tt -i,  -{}-cmb\_file\_raw=FILE cmb fits file of input data}}.
The freshly compiled version of the code did this correctly, providing the
user with a list of available options as expected.
\sloppy

\subsubsection{Scripts}

\fussy
My private notes of what were intended to record the most significant steps
taken in carrying out the project, along with more minor steps, were
used in the attempted reproduction of RBG08.  However, in trying to
reproduce these steps now, it is clear that the original notes were
not as complete and unambiguous as they should be.  For the purpose of
the current exercise in reproducibility, a completely fresh {\sc bash}
script was prepared, which verifies the sha512 checksums of input data
files and software packages that are downloaded from source rather
than provided within the security context of the host system. The
style of the new script is partly based on more recent attempts at
reproducibility in my own recent papers in which {\sc git} commit
hashes of the software\cite{Roukema17silvir} and a {\sc bash}
script for running the full software\cite{RO19flatness} were
provided. Some inspiration was taken from the {\sc make}-based reproducibility
framework\cite{Akhlaghi15} recently renamed
{\sc maneage}\cite{Akhlaghi2020}, but the structure of the
script is much less modular; it is a simple linear script with a
few minimal checks.

This situation illustrates the problem of ``insider knowledge'' being
required for the reproducibility of a paper, where ``insider'' also
includes knowledge that may still be coded in the scientist's brain,
but not in written form.

\subsubsection{Fortran 77--C interfacing}

A more fundamental problem in terms of coding and software environment
evolution is that this code uses a C front end and a Fortran 77
backend, configured and compiled together using {\sc autoconf} tools,
in a way that, to the best of my knowledge and that of my co-authors,
worked correctly in 2008. The key element for interfacing of Fortran
77 and C code that was recommended at the time was the use of {\sc
AC\_F77\_WRAPPERS} in the {\sc configure.ac}
file\cite{ACPROGF77}.  Three of the modular packages called by
the main code -- {\sc cosmdist}, {\sc astromisc} and {\sc isolat} --
also use {\sc AC\_F77\_WRAPPERS}.

The main files are {\tt circles\_f77.f}, 2023 lines long, named to
emphasise the expected obsolescence of the Fortran 77 language
standard; 21 Fortran 77 source files with a total of 11,616 lines of
code in {\tt lib/}; and 3415 lines of Fortran 77 code in the auxiliary
package {\tt astromisc/lib/}. For modernisation of this code, a
minimum approach would be to convert from Fortran 77 to Fortran 2008,
in which case Fortran--C compatibility conventions that are reasonably well
developed and implemented by the {\sc gcc/gfortran} family could be used.
Alternatively, {\sc f2c}, which continues to be available in distributions
such as Debian GNU/Linux, could be used to convert the Fortran code into C.
Any remaining bugs in the interfacing would be solvable within the standard
requirements of C, bypassing the issue of interlanguage communication.

For the purposes of this reproducibility test, the required re-coding
effort would be more than is presently justified. While this is, as
far as I know, the only free-licensed code for identified-circles
matching for the purposes of cosmic topology analysis for which
peer-reviewed research has been published, the techniques developed in
codes that are not publicly available under either free or non-free
licences have developed considerably since 2008.  Moreover, the
research field is extremely high risk: an observational confirmation
of the spatial topology of the Universe would be a historically
important discovery, but whether or not this measurement is feasible
remains highly speculative. To serve as a basis for long-term
projects in this particular field, the software would best be
rewritten according to modern standards of C and/or Fortran; and
the best tested, accurate, fast, well-coded free-licensed
auxiliary libraries, such as {\sc cgal} for geometrical purposes,
could be used to avoid ``reinventing the wheel''.

Given that a hurried attempt at a major refactoring of the code would
not only tend to extend beyond the scope of the aims of the ``Ten
Years Reproducibility Challenge'', it would also be pointless (a
systematic upgrade would better be done properly and thoroughly),
this reproducibility attempt was terminated, leaving it with
an untraced Fortran--C interfacing memory bug.
\sloppy

\section{Discussion}

Compiler evolution and Fortran/C compatibility are the main elements of
the difficulty in reproducing RBG08, together with the presence of some coding
that was not sufficiently robust to allow for the interface change.
In 2008, {\sc g77} was an obvious
choice of Fortran compiler in the stable distribution of Debian
GNU/Linux. Since the Debian community already had at the time a solid
reputation in terms of software security, verification of licensing
and fully transparent and participatory decision-making, this seemed
like a wise choice for reproducibility. Using a well-tested, widely used
compiler and code standard seemed like a better long-term sustainable choice
than using a compiler whose role was still being debated.
The main developer of {\sc g77} had already announced his intention to stop maintaining the
project in 2001\cite{Burley01g77}, but even by 2010, two years
after RBG08 was published, the community remained unclear regarding
the relationship between {\sc g95}, based on {\sc gcc}, versus
{\sc gfortran}, part of {\sc gcc}\cite{Bosscher10g95}.

However, also by 2010, {\sc gcc} already claimed to implement
compatibility between Fortran 2003 (ISO/IEC 1539-1:2004(E)) and ISO
C99 (ISO/IEC 9899:1999)\cite{GCCISOFortranC}, implemented at the
coding level by {\tt use ISO\_C\_BINDING} and {\tt bind(C, ...)}
declarations and the ability to declare C types in a Fortran module.

The IT group at the University of Oxford Department of Physics
recommend the modern interfacing methods, and describe
the {\sc g77} interfacing with C quite colourfully, stating that
``Part of the reason for the transition from g77 to gfortran is to
make mixing-in with C code simpler, and avoid (most of) the acts of
cruel and unusual programming which were previously required to get
the compilers' outputs to co-operate. Inevitably, the results of said
acts were almost inevitably fragile and non-portable.''\cite{Oxfordg77goneto}

An interesting question is whether a reproducibility framework such as
{\sc maneage}\cite{Akhlaghi2020}, which aims at a very high
standard of reproducibility with maximum modularity and minimal
dependencies, would have enabled easy reproducibility of the original
project.  The {\sc maneage} system would, in principle, have
configured, compiled, and installed all the original software
environment, including {\sc GSL} and {\sc g77} and a contemporaneous
version of {\sc gcc}, and would have encouraged the original project
to be modular enough with sufficiently many verification tests to
survive a decade of evolution of the software environment.  A key
question would be whether a more modern version of {\sc gcc} could
have compiled the old versions of {\sc g77} and {\sc gcc}.
It is quite realistic to expect that a {\sc maneage} reproduction
of RBG08 would succeed more than the {\sc bash} script method
provided in this project.

Would this project have been more easily reproducible had it been
written in a higher level language, such as python? The question
did not really arise at the time, since as is stated in the abstract
of RBG08, one of the key results was a speed-up in computation
time, a critical bottleneck for this type of research, in which
case the computational overhead of higher level languages renders
them impractical.

The aim of the {\em Ten Years' Reproducibility Challenge} is primarily
to get ``old code to run on modern hardware/software (with minimal
modifications)''.  Nevertheless, an interesting question would be how
much shifting back to an old software environment would be required to
run the existing code.  This raised a simple practical question: where can we
find an official, archived version of the source code of {\sc g77} in
the version likely to have been used when the paper was calculated and
published? This turned out to be non-trivial, because the nature of
the relation between {\sc g77} and {\sc gcc} was not sufficiently
known by me, nor was it easy to find using search engines.
The Debian GNU/Linux developers' guide to Fortran
updating from {\sc g77} to {\sc gfortran}, lasted edited in 2016,
states that ``g77 and g77-3.4 have been removed from the
archive''\cite{DebianGfortran}. It was only when the main developer
of {\sc g77}, James Craig Burley, kindly responded to a question posed
using a social networking feature provided on a git repository server,
that the packaging of {\sc g77} {\em inside} {\sc gcc} tarballs
became known to me.
Thus, while finding {\sc g77} source code was not as easy as expected,
once the detailed information about its relation with {\sc gcc} was known,
finding an archived version on a reputable webserver was found\footnote{\url{https://ftp.gnu.org/gnu/gcc/gcc-3.4.6/}}.
However, attempts at compiling
{\sc g77-3.4.6} within {\sc gcc-3.4.6} on Debian GNU/Linux stable (9.12)
were unsuccessful.

\section{Conclusion}

\fussy
It is ironical that in the field of cosmic topology, not only are most
software packages only available privately with unknown licences (as
tends to be the case in astronomy as recently as
2015\cite{Allen18}), but the code that is explicitly
free-licensed, publicly distributed and having peer-reviewed published
results has turned out to be less easy to reproduce than
expected. This is, unfortunately, consistent with typical reports on
science research paper
reproducibility\cite{Ioannidis2009,Chang15,Stodden18}.  The
specific bottleneck suspected of leading to memory errors in this case
was that the effort required to update the Fortran 77 files at the
heart of the code, interfaced with a C front end, and compiled with
the current {\sc gfortran} compiler from within {\sc gcc} rather than
with the older, discontinued\cite{Burley01g77} {\sc g77}
compiler, risked being too great to be justified on any short time
scale. While Fortran has remained actively used by scientists since
more than half a century ago and in its modern standards continues to
be used actively, and the original {\sc circles} package was prepared
using the powerful GNU {\sc autotools}, a robust interface and
standards for compiling C and Fortran code together have only evolved
quite recently\cite{GCCISOFortranC}.

While the results of the paper are not as trivially reproducible as
they appeared to be, the requirement of the Ten Years Challenge for
the code to be placed in an online git repository, which in this case
is \url{https://codeberg.org/boud/0807.4260}, resulted in confirmation
that the source code is fully free-licensed, including all libraries
and other auxiliary software packages, and the input data files remain
publicly available online. Refactoring the code into a format such as
{\sc maneage}\cite{Akhlaghi15,Akhlaghi2020} would be a potential
way forward of shifting the research field towards a more
completely \emph{open-science} phase. Anyone interested in modernising
the software and completing the reproduction of the original results
is welcome to contact the author of this paper.

\section*{Acknowledgments}
{\footnotesize Thank you to Konrad Hinsen (the reviewer) and James
  Craig Burley for several helpful comments.
  Part of this research has been supported by the ``A
  next-generation worldwide quantum sensor network with optical atomic
  clocks'' project of the TEAM IV programme of the Foundation for
  Polish Science co-financed by the European Union under the European
  Regional Development Fund.  Part of this research has been supported
  by the Polish MNiSW grant DIR/WK/2018/12.  Part of this research has
  been supported by the Pozna\'n Supercomputing and Networking Center
  (PSNC) computational grant 314.\sloppy}
 
\hypersetup{linkcolor=black,urlcolor=darkgray}
\renewcommand\emph[1]{{\bfseries #1}}
\bibliographystyle{mnras_openaccess_note}

\end{document}